\documentclass[a4paper,amsmath,amssymb,superscriptaddress,aps,prb,reprint,showpacs]{revtex4-2} 
\usepackage[caption=false]{subfig}
\usepackage{xcolor,graphicx} 
\usepackage{MnSymbol} 
\usepackage{mathtools}
\usepackage{braket} 
\usepackage{tikz} 

\usepackage[utf8]{inputenc} 
\usepackage{hyperref}
\usepackage{here}
\usepackage{bbold}
\usepackage[version=4]{mhchem}

\begin{document}

\title{Robustness of topological edge states in alternating spin chains against environment}

\author{Alexander Sattler}
\author{Maria Daghofer}
\affiliation{
Institut f\"ur Funktionelle Materie und Quantentechnologien,
Universit\"at Stuttgart,
70550 Stuttgart,
Germany}

\date{\today}

\begin{abstract}
  Both the Haldane spin-$1$ chain and dimerized chains of
  spin-$1/2$  exhibit topologically protected edge states that
  are robust against specific perturbations.  Recently, such spin
  chains have been specifically assembled on surfaces and we
  investigate here the robustness of these edge states against
  coupling to the surface. Since no physical system can be considered perfectly isolated, it
  is crucial to examine whether topological robustness is maintained in the
  presence of environmental coupling.
  We apply exact diagonalization to a Lindblad master equation that couples an
  alternating Heisenberg spin chain based on spins $1/2$ to a surface via
  various jump operators. The robustness of topological states is assessed via the
  time evolution of quantities such as the ground-state degeneracy, correlation
  function, entropy, and magnetization of edge states.
  We investigate chains built from dimers with antiferromagnetic and
  ferromagnetic intra-dimer coupling, which resemble
  Su-Schrieffer-Heeger and the Haldane models, resp., and assess the
  impact of $z$-axis anisotropy and longer-ranged couplings. Generally,
  we find that signatures of  topological properties are 
  more robust in Su-Schrieffer-Heeger-like chains than in
  Haldane-like chains.
\end{abstract}

\maketitle

\section{Introduction} \label{sec:introduction}

Extensive research is currently centered on topological order and symmetry-protected topological order (SPT), alongside their associated characteristics, spanning various systems and dimensions~\cite{RevModPhys.89.040502,RevModPhys.89.041004,RevModPhys.88.035005,RevModPhys.83.1057,RevModPhys.82.3045}. 
These states transition to topologically trivial ones by closing a gap, emphasizing the importance of edge states in identifying topologically nontrivial states. 
There are ongoing efforts to identify and explore such states, regardless of whether they arise naturally in solids or artificial quantum systems.

The study of closed topological systems, particularly focusing on their edge-state properties, is well established.
However, it is crucial to consider interactions with the environment,
given that no physical system is perfectly isolated.  
The impacts of these interactions, including phenomena like decoherence~\cite{Entanglement_and_Decoherence,breuer_petruccione_book},  remain relatively unexplored. 
Even topological systems cannot be expected to be protected from decoherence, raising
questions about the stability of  SPT and topological order in the presence of
environmental effects.

For any practical use of topological properties, e.g., in topological quantum
computers~\cite{RevModPhys.88.041001,doi:10.1126/science.270.5242.1633,PhysRevLett.85.1762,10.1002/1521-3978,Ladd2010,PhysRevA.51.1015,10.21468/SciPostPhys.3.3.021,RevModPhys.80.1083,Andrew_Steane_1998},
it would be essential to know how far the topological stability observed in closed systems
persists when it is coupled to an environment. 
In particular, there are theoretical proposals that suggest the use of edge
states of the Haldane chain for quantum
computations~\cite{edge_quant_commun,RevModPhys.80.1083,015001,app9030474,Jaworowski2017,PhysRevLett.105.110502,PhysRevLett.105.040501,PhysRevA.86.032328,app9030474,Hsieh_2012,PhysRevB.104.125402}
and ongoing discussion explores their potential applications in
spintronics~\cite{doi:10.1126/science.1065389,HIROHATA2020166711,RevModPhys.76.323,doi:10.1146/annurev-conmatphys-070909-104123}.

The effect of the environment on topological edge states demonstrates a
complex
interplay~\cite{Klett2018,PhysRevA.98.013628,PhysRevB.109.035114,PhysRevB.85.121405,PhysRevB.86.155140,PhysRevB.104.094306,PhysRevB.108.155123,PhysRevLett.127.086801,PhysRevLett.126.237201,PhysRevB.94.201105,PhysRevB.88.155141,Diehl2011,PhysRevLett.109.130402,10.1088/1367-2630/15/8/085001}.   
While most results indicate a lack of topological robustness, some suggest a
certain resilience to noise or even the emergence of novel phenomena that do not occur in
closed systems, such as topologically nontrivial steady
states~\cite{PhysRevA.91.042117,Verstraete2009,dai2023steadystate,Diehl2011}.  
Experimentally, edge states of a chain of superconducting
qubits have been shown to be somewhat robust against noise~\cite{doi:10.1126/science.abq5769}.

A well-known and expensively studied example of a state with SPT order with
edge states is the Haldane phase~\cite{PhysRevB.48.3844,PhysRevB.81.064439,
  PhysRevB.80.155131,RevModPhys.89.040502,RevModPhys.89.041004,I_Affleck_1989,inbook_regnault,PhysRevB.85.075125}. It is based on the conjecture of
Haldane~\cite{PhysRevA.93.464,PhysRevLett.50.1153} and represents the ground
state of a spin-$1$ chain featuring antiferromagnetic (AFM) coupling between
spins. 
This ground state can be comprehended by dividing each spin-$1$ into two
spin-$1/2$, yielding a spin-$1/2$ chain with alternating coupling constants.  
The formation of singlet
states~\cite{PhysRevB.85.075125,inbook_regnault,I_Affleck_1989,RevModPhys.89.041004,PhysRevB.48.9555,PhysRevB.46.8268,PhysRevB.45.2207,PhysRevB.46.3486,PhysRevLett.65.3181,1904.02102}
results in residual spin-$1/2$ edge states at both ends, which are referred to
as topologically protected edge states.  
These edge spins can form either a global singlet or triplet, but as they
decouple for longer chains, all four states are degenerate.
The topological energy gap above this ground-state manifold can be measured  by neutron
scattering, susceptibility measurements, and magnetization measurements~\cite{PhysRevB.54.R6827,PhysRevB.39.4820,10.1063/1.340736,PhysRevLett.56.371,J_P_Renard_1987,PhysRevLett.63.86,PhysRevB.38.543,REGNAULT199571,DARRIET1993409,PhysRevLett.63.1424,Cizmar_2008,PhysRevLett.90.087202}.
Furthermore, the presence and behavior of edge states can be probed using techniques such as electron spin
resonance~\cite{PhysRevLett.65.3181,PhysRevLett.67.1614,doi:10.1143/JPSJ.67.2514,Cizmar_2008,PhysRevLett.95.117202}, nuclear magnetic resonance~\cite{PhysRevLett.83.412} and inelastic neutron scattering~\cite{PhysRevLett.90.087202}.

The Haldane phase can be realized in artificial quantum systems.
It has been theoretically~\cite{PhysRevLett.111.167201} shown that edge states
of Haldane chains are accessible to a scanning tunneling microscope (STM) and
they have been measured for both spin-$1$ chains~\cite{Mishra2021,Zhao2023}
and alternating spin-$1/2$ chains~\cite{2403.14145,zhao2024tunable}.  
Furthermore, it was proposed that a Haldane chain may be realized in various
systems based on quantum
dots~\cite{Hsieh_2012,PhysRevB.104.125402,app9030474,2404.12207,SHIM20102065,Jaworowski2017,PhysRevB.109.085112}.

We previously discussed the properties of the topological edge states for
the closed alternating spin-$1/2$ chain  in
Ref.~\cite{sattler2024topologically}. Our main finding had been that
topological properties are easier to observe in chains that have AFM coupling
throughout -- and thus resemble an interacting Su-Schrieffer-Heeger
(SSH) model --
than they are in Haldane-like systems. Here, we aim to continue this work by
incorporating the coupling with the surface using the Lindblad master equation
(LME)~\cite{breuer_petruccione_book}.

In Sec.~\ref{section:model} we present the Hamiltonian of our model and introduce the LME.
In Sec.~\ref{sec:phase_diagram} we analyze the robustness of the ground-state degeneracy based on low energy gaps. 
To examine the behavior of edge states, we explore the time evolution of the edge-state magnetization in Sec.~\ref{sec:S_z_exp_val} and the correlation functions in Sec.~\ref{sec:correlation}. 
Furthermore, different criteria are discussed to analyze the time evolution of the chain, including entropy, fidelity, and purity in Sec.~\ref{sec:entropy} and the spectral gap in Sec.~\ref{sec:spectral_gap}.
Finally, Sec.~\ref{sec:discuss} provides a summary of our findings as
well as an outlook to promising future studies. 
 
\section{Model and Methods}\label{section:model}

We investigate alternating Heisenberg spin-$1/2$ chains~\cite{Chen_2019,doi:10.1143/JPSJ.62.1463,PhysRevB.59.11384,PhysRevLett.89.077204,PhysRevB.74.144414,doi:10.1143/JPSJ.62.3357,Bahovadinov_2019,Haghshenas_2014,PhysRevB.46.8268,PhysRevB.46.3486,10.1088/0953-8984/26/27/276002,doi:10.1143/JPSJ.58.4367,PhysRevB.46.3486,PhysRevB.45.2207,PhysRevB.48.9555,10.7566/JPSJ.85.124712,1508.06129,10.1088/0953-8984/27/16/165602,PhysRevB.87.054402,1212.6012,1904.02102,PhysRev.165.647,PhysRevB.46.3486,PhysRevB.46.8268,PhysRevB.45.2207,PhysRevB.63.144428,doi:10.1143/JPSJ.61.1879} with nearest-neighbor (NN) couplings 
\begin{equation}
\begin{aligned}
H_\mathrm{NN} &= J_1 \sum_{i=1}^{N/2}   \left(  S_{2i-1}^{x} S_{2i}^{x} 
+  S_{2i-1}^{y} S_{2i}^{y}
+ \Delta_z S_{2i-1}^{z} S_{2i}^{z} \right) \\ 
&+ J_2 \sum_{i=1}^{N/2-1}   \left(  S_{2i}^{x} S_{2i+1}^{x} 
+  S_{2i}^{y} S_{2i+1}^{y}+ \Delta_z S_{2i}^{z} S_{2i+1}^{z} \right), 
\label{eq:Hamilton_NN}
\end{aligned}
\end{equation}
where $J_1$ and $J_2$ are the alternating coupling constants  and $\Delta_z$ introduces $z$-axis anisotropy. 
$\Delta_z \gg 1$ leads to more Ising-like spins, while $\Delta_z \ll 1$ would imply $x$-$y$ anisotropy. 
We restrict ourselves to chains with an even number of spins $N$.
Topological edge states emerge only with an AFM~\cite{PhysRevB.87.054402} coupling $J_2 > 0$, where $J_2$ is set as an energy unit, i.e., $J_2=1$ and $\hbar =1$. 
Open boundary conditions (OBC) are employed for the examination of edge states unless explicitly specified that periodic boundary conditions (PBC) are applied.

Additionally, we include next-nearest-neighbor (NNN) coupling~\cite{1508.06129,1212.6012,10.7566/JPSJ.85.124712,1904.02102,10.1088/0953-8984/26/27/276002,PhysRevB.109.094439,Chatterjee_2024} 
\begin{align}
H_{\mathrm{NNN}} =  J_\mathrm{NNN} \sum_{i=1}^{N-2}\left( S_{i}^{x} S_{i+2}^{x} 
+  S_{i}^{y} S_{i+2}^{y} + \Delta_z S_{i}^{z} S_{i+2}^{z} \right).
\label{eq:Hamilton_NNN}
\end{align}
Thus, the complete Hamiltonian is $H = H_\mathrm{NN} + H_\mathrm{NNN}$, and we provide a detailed discussion of this Hamiltonian in Ref.~\cite{sattler2024topologically}.

Given our interest in spin chains on surfaces, we need to account for
interactions between the chain and the surface.  
Our approach here is to use an effective model that treats the spin
chain as an open quantum system, with the environment being the
surface.  
Open quantum systems can be described by master equations~\cite{breuer_petruccione_book}.
Lindblad~\cite{lindblad_derivation}
and Gorini, Kossakowski and Sudarshan~\cite{doi:10.1063/1.522979} derived the general form of a quantum dynamical semigroup generator $\mathcal{L}$  that describes Markovian dynamics of density matrices. 
The LME can be written as
\begin{align}
\dot{\rho} = -\mathrm{i} [H,\rho] +  \gamma  \sum_i  \left( L_i^{\vphantom{\dagger}} \rho L_i^{\dagger} -\frac{1}{2} \Big\{\ L_i^{\dagger} L_i^{\vphantom{\dagger}}, \rho \Big\}\   \right),
\label{eq:lindblad_LME}
\end{align}
where $L_i$ are the jump operators, $\gamma>0$  the dissipation strength, and $\hbar=1$.  
We adopt the initial condition that the coupling to the environment is switched on at $t=0$.
The derivation of the LME  can be found in Refs.~\cite{Entanglement_and_Decoherence,PhysRevB.104.094306,breuer_petruccione_book,lindblad_derivation,doi:10.1063/1.522979, brasil.fanchini.napolitano.2013,1902.00967,10.1063/1.5115323}.
The details of the numerical methods are discussed in  Appendix~\ref{appendix:LME}.

The results of the LME of course depend on the choice of jump operators.
We treat $L_i^{\vphantom{z}}=S_i^z$ and  $L_i^{\vphantom{x}}=S_i^x$.
$L_i^{\vphantom{z}}=S_i^z$ models an interaction known as dephasing noise, which causes  decoherence~\cite{breuer_petruccione_book,PhysRevE.92.042143,PhysRevLett.111.150403,PhysRevA.97.052106,PhysRevB.99.174303,PhysRevE.83.011108},
see also Appendix~\ref{sec:appendix_L_z}. The summation over the jump
operators in Eq.~\eqref{eq:lindblad_LME} corresponds to a sum over all sites of the chain
for either $L_i^{\vphantom{z}}=S_i^z$ or $L_i^{\vphantom{x}}=S_i^x$,
the two scenarios are denoted by $L_z$ and $L_x$.

Instead of $L_i^{\vphantom{x}}=S_i^x$, one could also consider the
raising and lowering operators $L_{i,1}^{\vphantom{+}}=S_i^+$ and
$L_{i,2}^{\vphantom{-}}=S_i^-$ as  jump operators, which can be used to model transitions
between two energy levels, such as those resulting from spontaneous
emission or absorption.
Compared to the combined $S_i^x$, a LME with raising and lowering
operators mixes  elements of the density matrix more strongly, see also
Appendix~\ref{sec:appendix_S_+_S_-}.
The jump operators $L_i^{\vphantom{x}}=S_i^x$ chosen here have their own
physical interpretation, namely, random interactions with a noisy
environment~\cite{PhysRevB.94.134107}.

We utilize the LME to calculate the time evolution of the eigenstates of the Hamiltonian. 
Writing the Hamiltonian in matrix form necessitates the selection of a
basis, which involves choosing a quantization axis for the spin. In
this work, we adopt the commonly used $z$-direction as the
quantization axis. 
Since the eigenstates of the Hamiltonian share the same quantization
axis as the Hamiltonian itself, the initial density matrices used in
the LME also have this quantization axis. This explains the observed
difference between $L_x$ and $L_z$ in the case of $\Delta_z = 1$,
which is evident, for example, in the entropy.

\section{Results} \label{sec:results}

\subsection{Ground-state degeneracy}  \label{sec:phase_diagram}

In Ref.~\cite{sattler2024topologically} we previously examined the utility of an energy gap ratio, denoted as $\Theta_\mathrm{PT}$,  as a criterion to delineate regions where it is likely that topological edge states in closed chains can be observed.
Here, we aim to extend this analysis by utilizing this ratio to assess the longevity of edge states in the presence of surface coupling.

To achieve this, we need to adapt the definition of
$\Theta_\mathrm{PT}$, because energy crossings during the time
evolution lead to changes in the ranking of energy states.  
Consequently, the definition of $\Theta_\mathrm{PT}$, which only takes
into account the five lowest-energy states at $t = 0$, becomes
invalid for $t > 0$. However, calculating time evolution for
\emph{all} eigenstates would be impractical. Fortunately, the
topological energy gap closes rapidly, at time scales where higher-energy states
generally remain well above the lowest energy. 
Therefore, focusing on lower-energy states usually continues to be
sufficient, with the $70$ lowest states generally yielding reliable
results, and more states are only needed in rare cases.
Before calculating the ratio of energy gaps, one, therefore, has to keep
track of any energy crossings at each time step, which
leads to the modified definition: 
\begin{equation}
\begin{aligned}
\alpha(t) &= \max\limits_{\substack{i=1,2,3 }} \frac{\Delta E_i(t)}{\Delta E_{4-5}(t)} \\
\Theta_\mathrm{PT}(t) &\coloneqq
\left\{
\begin{array}{ll}
\alpha (t)
 \hspace{0.1cm} &\mathrm{if} \left(\max\limits_{\substack{j=1,2,3,4 }} E_j(t)\right)<\left(\min\limits_{\substack{j\geq 5 }} E_j(t)\right)  \\
1 &\mathrm{else}. \\
\end{array}
\right. 
\label{eq:theta_def_open}
\end{aligned}
\end{equation}
Here, $E_i$ denotes the energies ordered by ascending energy levels in the closed system, $\Delta E_i(t)$ the $i$th lowest gap,
and $\Delta E_{4-5}$ represents the energy gap between the fourth and fifth lowest energy levels.

Next, we establish a threshold $\Theta_\mathrm{PT}^\mathrm{tran}$
beyond which edge states observation is probably viable.  
Although the specific value of $\Theta_\mathrm{PT}^\mathrm{tran}$ is
debatable, the main outcomes remain robust across variations.  
We adopt $\Theta_\mathrm{PT}^\mathrm{tran}=0.5$ from our previous
discussions about closed   chains~\cite{sattler2024topologically}. 

\begin{figure}
  \includegraphics[width=\columnwidth]{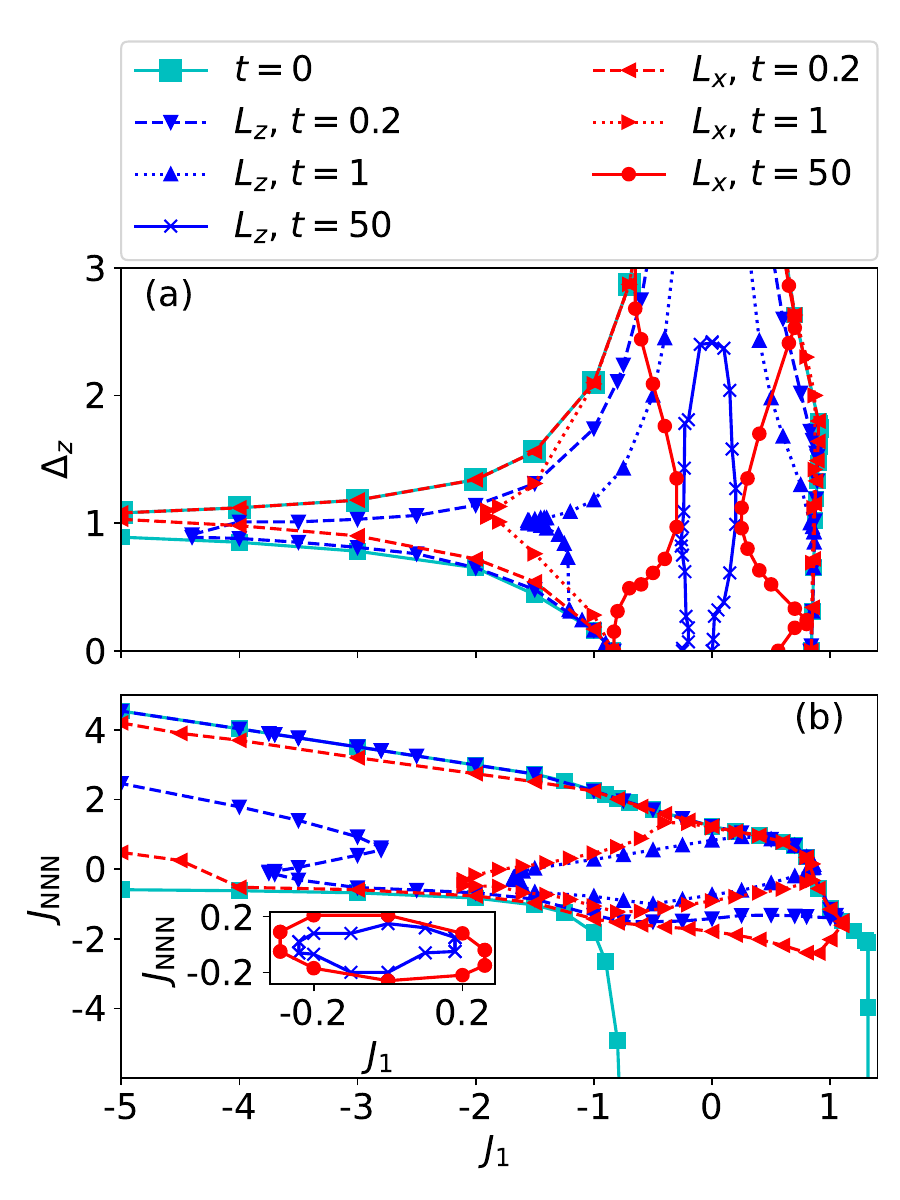}
   \caption{ $\Theta_\mathrm{PT}$-diagram that shows the set of
     parameters where
     $\Theta_\mathrm{PT}(t)<\Theta_\mathrm{PT}^\mathrm{tran}$ for  NNN
     coupling and $z$-anisotropy influence on the time evolution for
     $L_z$ and $L_x$. The chain parameters are set to $N=8$,
     $\gamma=1$, with two scenarios: (a) $J_\mathrm{NNN}=0$, and (b)
     $\Delta_z=1.001$ for $L_z$, and   $\Delta_z=1$  for $L_x$.  
\label{fig:N_8_phase_diagram}}
\end{figure}

\begin{figure}
  \includegraphics[width=1.15\columnwidth]{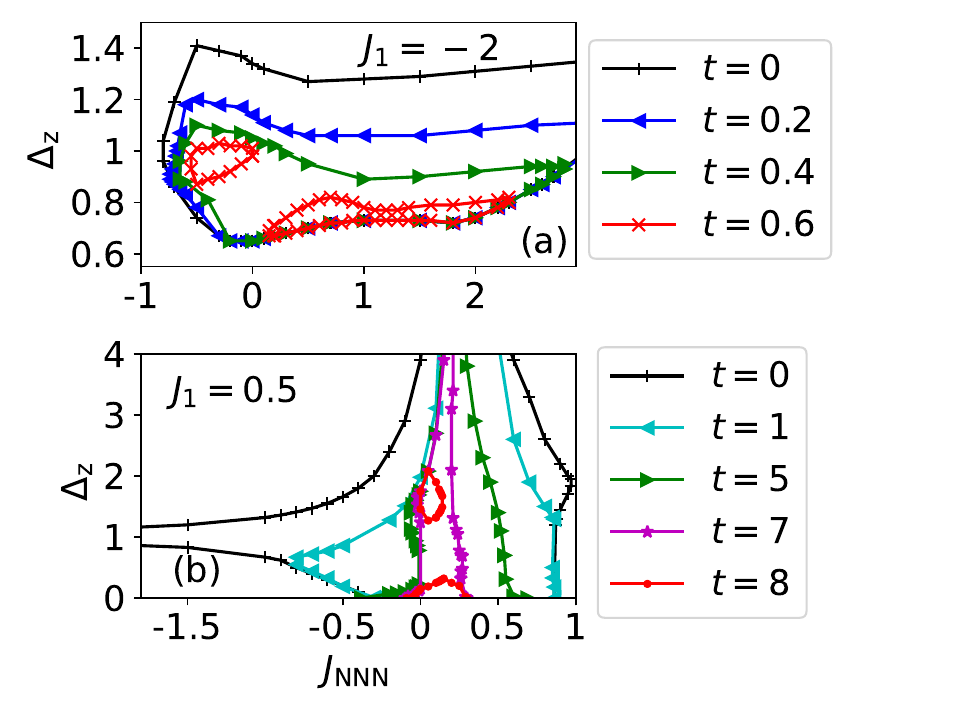}
   \caption{$\Theta_\mathrm{PT}$-diagram that shows the set of parameters where $\Theta_\mathrm{PT}(t)<\Theta_\mathrm{PT}^\mathrm{tran}$. Impact of the combination of NNN coupling and $z$-anisotropy on $\Theta_\mathrm{PT}(t)$ for a chain with $L_z$, $N=8$,  $\gamma=1$: (a) $J_1=-2$ and    (b) $J_1=0.5$.
\label{fig:L_z__z_anis__NNN__phase_diagram}}
\end{figure}

The diagrams presented in this section illustrate parameter sets where  $\Theta_\mathrm{PT}(t)<\Theta_\mathrm{PT}^\mathrm{tran}$. 
However, it is important to emphasize that these diagrams are not topological phase diagrams, as $\Theta_\mathrm{PT}(t)$ does not serve as a sufficient criterion for a topological phase transition. 
To be more precise, $\Theta_\mathrm{PT}(t)$ serves as a criterion for determining the lifetime of the quasi-fourfold degenerate ground state.

In certain scenarios, the ground-state degeneracy $D_\mathrm{GS}$ can serve as a characterization tool for topological phases~\cite{RevModPhys.89.041004,PhysRevB.41.9377,A_Short_Course_on_Topological_Insulators,CHEN2020126168,PhysRevB.89.075121,1904.02102}. 
However, $D_\mathrm{GS}$ alone is inadequate as a topological criterion, especially for the Haldane phase~\cite{PhysRevB.81.064439,PhysRevB.80.155131}.
Additionally, even the often used string order parameter~\cite{PhysRevB.40.4709,PhysRevLett.66.798} is not a sufficient criterion~\cite{PhysRevB.81.064439,PhysRevB.80.155131}.
The Haldane phase, as an SPT phase, is stable, as long as the
protecting symmetries are
preserved~\cite{PhysRevB.81.064439,PhysRevB.80.155131}.  
Thus it would be necessary to discuss the symmetries for creating
phase diagrams. 
Instead of constructing topological phase diagrams, our focus here is
to identify parameters conducive to edge state observation in
experiments. Consequently, this section primarily discusses
$D_\mathrm{GS}$.

\begin{figure}
  \includegraphics[width=\columnwidth]{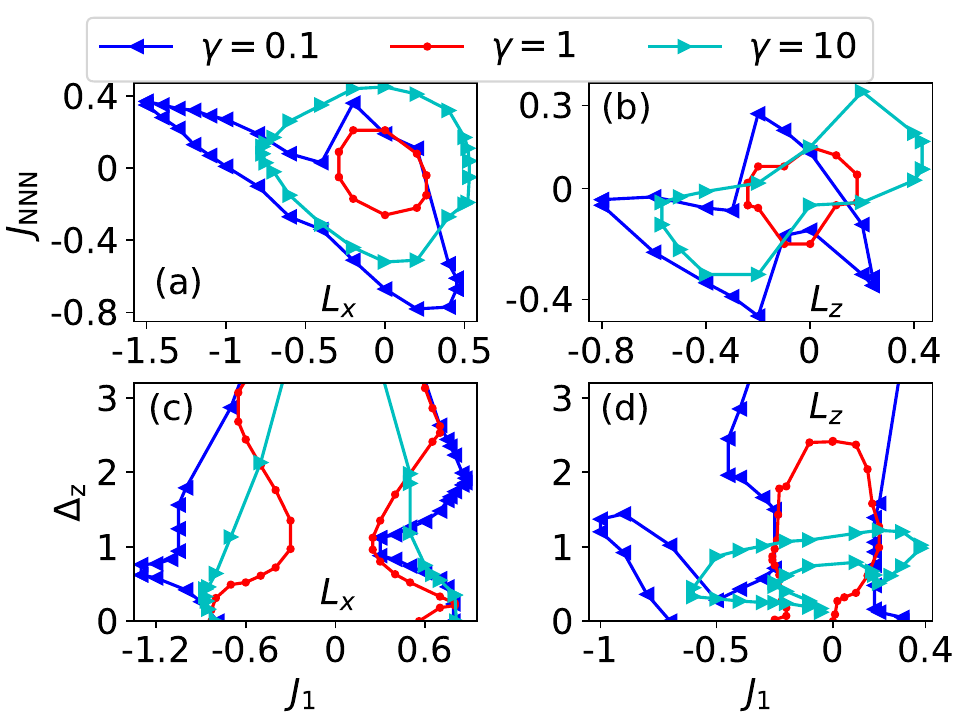}
   \caption{$\Theta_\mathrm{PT}$-diagram that shows the set of parameters where $\Theta_\mathrm{PT}(t)<\Theta_\mathrm{PT}^\mathrm{tran}$. The parameters are $t=50$, $N=8$,  $L_x$ in (a,c), $L_z$ in (b,d) and  $J_\mathrm{NNN}=0$ (c,d). There is $\Delta_z=1$ in (a) and in (b) there is $\Delta_z=1.001$ for $\gamma=1,10$ and $\Delta_z=1$ for $\gamma=0.1$.  
\label{fig:time_50_phase_diagram_gamma}}
\end{figure}

Haldane-like chains, i.e., chains resembling the Haldane spin-$1$ AFM
chain ($J_1 < -1$), lose their $D_\mathrm{GS}$ rapidly. This contrasts
with chains from the dimer scenario ($|J_1| < 1$, essentially a spin
variant of the Su-Schrieffer-Heeger model), which maintain
$\Theta_\mathrm{PT}(t) < \Theta_\mathrm{PT}^\mathrm{tran}$ for longer
times. This is shown in Fig.~\ref{fig:N_8_phase_diagram}, 
where the time evolution of the $\Theta_\mathrm{PT}$-diagrams for
$z$-anisotropy (see Eq.~\eqref{eq:Hamilton_NN}) and NNN coupling (see  Eq.~\eqref{eq:Hamilton_NNN}) are depicted. 
Additionally, Fig.~\ref{fig:time_50_phase_diagram_gamma} illustrates
the impact of $\gamma$ on the $\Theta_\mathrm{PT}$-diagram,
showing that this holds true for all values of $\gamma$. 
Finally, the more robust ground-state degeneracy of the SSH-like
scenario aligns perfectly with the results observed in closed
chains~\cite{sattler2024topologically}.

Furthermore, we observe in Fig.~\ref{fig:N_8_phase_diagram} similar
trends in the time evolution for different jump operators, $L_x$ and
$L_z$. Generally, the region where
$\Theta_\mathrm{PT}(t)<\Theta_\mathrm{PT}^\mathrm{tran}$  is larger
for $L_x$ than for $L_z$ but the fundamental trend persists: $D_\mathrm{GS}$ of
chains from the topological dimer scenario is more robust than that of
Haldane-like chains. 
Intriguingly, the
$D_\mathrm{GS}$  of chains from the dimer scenario exhibits remarkable
robustness even at rather large $z$-anisotropies, see Fig.~\ref{fig:N_8_phase_diagram}(a).

The impact of  NNN coupling $J_\mathrm{NNN}$ is presented in 
Fig.~\ref{fig:N_8_phase_diagram}(b). Again, $D_\mathrm{GS}$ of Haldane-like chains remains less
robust over time compared to that of chains from the topological dimer
scenario. However, we also find a striking difference to closed-chain
results reported in Ref.~\cite{sattler2024topologically}, where FM $J_\mathrm{NNN}<0$ was found to
support GS degeneracy. This is no longer the case in open chains,
where GS degeneracy is lost fast.

Let us note that the $\Theta_\mathrm{PT}$-curves for $L_z$ in
Figs.~\ref{fig:N_8_phase_diagram}(b) and~\ref{fig:time_50_phase_diagram_gamma}
were obtained using a tiny $z$-axis anisotropy $\Delta_z =1.001$.
This physically irrelevant deviation from $\Delta_z =1$ ($\Delta_z
=0.999$ also works), was introduced to remove rather chaotic behavior
that were found for $\Delta_z =1$ and $\gamma =1, 10$. Similarly
erratic behavior was not observed for $L_x$, and it might be due to
numerics or finite-size effects.

Figure~\ref{fig:L_z__z_anis__NNN__phase_diagram} shows
$\Theta_\mathrm{PT}$-diagrams for chains with NNN coupling and
$z$-anisotropy for $L_z$.   
 Similar to Fig.~\ref{fig:N_8_phase_diagram}, the $D_\mathrm{GS}$ of
 chains from the dimer scenario  exhibit greater robustness compared
 to  Haldane-like chains. 
Depending on the NN-coupling, frustrated chains can exhibit a more
(e.g., $J_1=0.5$) or less (e.g., $J_1=-2$) robust $D_\mathrm{GS}$ than
non-frustrated chains. 
Combinations of $\Delta_z$ and $J_\mathrm{NNN}$ can increase the
robustness compared to isotropic chains lacking NNN coupling.

In the time evolution of $\Theta_\mathrm{PT}(t)$, the quantum Zeno
effect (QZE)~\cite{breuer_petruccione_book,10.1063/1.523304} occurs in
some cases, but not in others. 
For $\gamma \ll 1$, robustness consistently decreases with increasing
$\gamma$, whereas for $\gamma \gg 1$, it may either increase or
decrease, as shown in Fig.~\ref{fig:time_50_phase_diagram_gamma}. This
suggests that, in certain cases, robustness exhibits a minimum as a
function of $\gamma$. The QZE is generally more likely to emerge in
dimerized chains than in Haldane-like chains, though exceptions
exist. Moreover, its occurrence depends on the choice of jump
operators and all Hamiltonian parameters. 
In cases where the QZE is observed, the minimum in robustness occurs
approximately at $\gamma \sim J_\mathrm{NN}$, with deviations becoming
more pronounced with increasing disparity between $J_1$ and $J_2$.

\begin{figure}
  \includegraphics[width=\columnwidth]{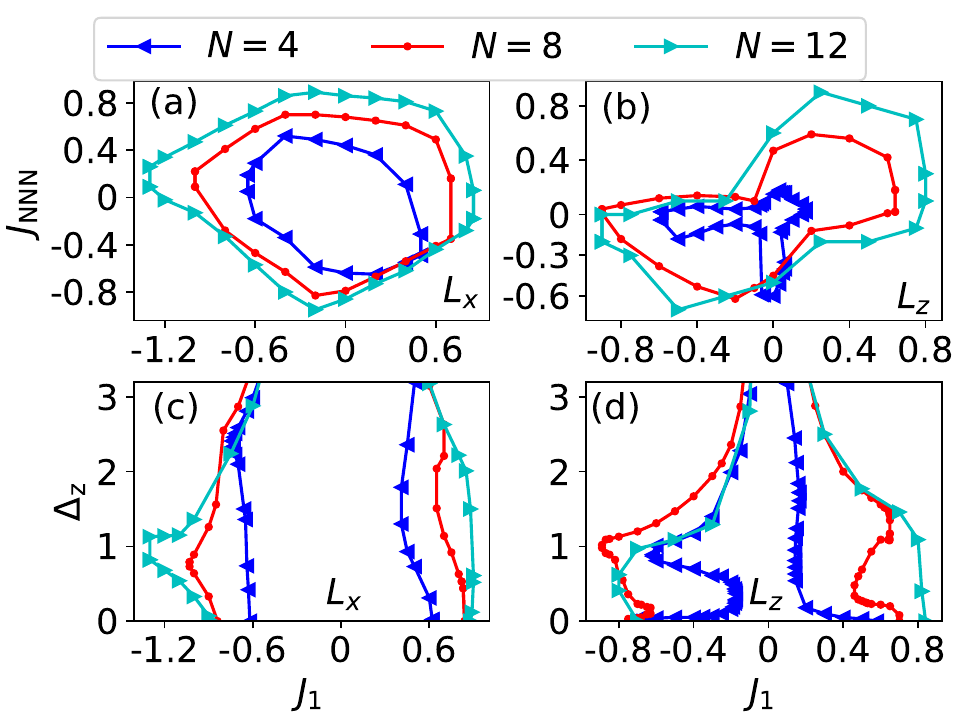}
   \caption{$\Theta_\mathrm{PT}$-diagram that shows the set of parameters where $\Theta_\mathrm{PT}<\Theta_\mathrm{PT}^\mathrm{tran}$. Finite-size effects for chains with $t=5$, $\gamma=1$, and  $L_x$ for (a,c),  $L_z$ for (b,d) and $\Delta_z=1$ (a,b) and $J_\mathrm{NNN}=0$ for (c,d).
\label{fig:time_5_phase_diagram_finite_size}}
\end{figure}

Apart from the aforementioned effects, the impact of $\gamma$ on the
time evolution of the $\Theta_\mathrm{PT}$-diagrams in
Figs.~\ref{fig:N_8_phase_diagram},
~\ref{fig:L_z__z_anis__NNN__phase_diagram}
and~\ref{fig:time_5_phase_diagram_finite_size} primarily appears in
differences in the time labels within the legend, while the overall
shape of the diagrams remains largely consistent.

The primary finite-size effect,  as shown in
Fig.~\ref{fig:time_5_phase_diagram_finite_size},   is that as $N$
increases, robustness generally increases, though there are some
exceptions.  
However, even for the smallest possible chain length ($N=4$), a
significant parameter space remains that allows for the observation of
edge states.  

\subsection{Edge-state magnetization}\label{sec:S_z_exp_val}

We look here at time- and site-dependent expectation values
$\langle S_i^z \rangle(t)$,  where $i$ denotes the site. 
Before discussing the short-term time evolution, we will first look at
the long-term time evolution, i.e., the steady state (ss): 
\begin{align}
\langle S_i^z \rangle_\mathrm{ss}=
\left\{
\begin{matrix}
0 \hspace{0.5cm} &\mathrm{for} \hspace{0.5cm}  L_x \ \\
\frac{1}{N}\langle S_\mathrm{chain}^z \rangle \hspace{0.5cm} &\mathrm{for} \hspace{0.5cm}  L_z.
\end{matrix}
\right. 
\label{eq:s_z_steady_state}
\end{align} 
This steady-state property is independent of the initial state of the
chain and of the observed site, making it also independent of the topology. 
The steady-state value for $L_z$ can be explained by the conservation
of $\langle S_\mathrm{chain}^z\rangle$~\cite{PhysRevA.97.052106,PhysRevE.92.042143}. 
The degeneracy of the steady state is determined by the jump operator. 
For $L_z$ the degeneracy of the steady state is $N+1$.
This can be easily understood because $L_z$ conserves $\langle
S_\mathrm{chain}^z\rangle$, so that each allowed value of $ \langle
S_\mathrm{chain}^z \rangle $ contributes one steady state.  
The degeneracy of the steady state for $L_x$ is also $N+1$ for
isotropic chains, and two if the chain is anisotropic. 
The steady state for $L_x$  is approached via
 \begin{align}
\langle S_\mathrm{chain}^z\rangle(t) = \langle S_\mathrm{chain}^z\rangle(0)\  \mathrm{e}^{-\frac{\gamma t}{2}}.
\label{eq:s_z_chain_l_x}
\end{align}
This exponential decay is identical to that of a single spin-$1/2$ with $S^x$ as the jump operator
\begin{align}
\langle S^z \rangle(t)=\langle S^z\rangle(0)\ \mathrm{e}^{-\frac{\gamma t}{2}}.
\end{align}
The derivation for a single spin is shown in Appendix~\ref{sec:appendix_L_x}.

We then look at the time evolution towards the steady-state values,
starting from edge-state spins of the two states within the 
ground-state manifold that have $|\langle
S_\mathrm{chain}^z\rangle(0)|=1$, i.e., two of the triplet states.
While
\begin{align}
 \langle  S_\mathrm{chain}^z \rangle = \sum_i \langle  S_i^z \rangle
\end{align}
denotes  the magnetization of the total chain, it is carried by the
two edge spins for topological chains, i.e., $\langle
S_\mathrm{edge}^z \rangle=\pm 1/2$.

\begin{figure}
  \includegraphics[width=\columnwidth]{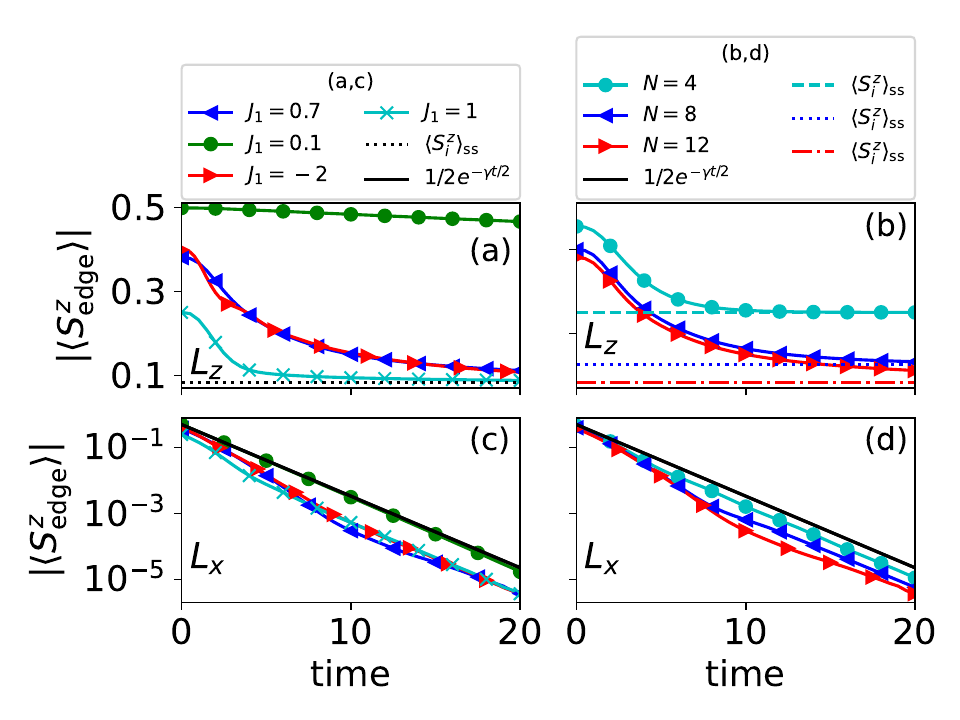}
    \caption{$\langle S_\mathrm{edge}^z\rangle(t)$ depending on $J_1$ and the number of spins $N$ for the triplet states with $\langle S_\mathrm{chain}^z\rangle(0)=\pm1$  for  chains with $\gamma=1$, $J_\mathrm{NNN}=0$ and $\Delta_z=1$. The remaining parameters are  $L_z$ (a,b) and  $L_x$ (c,d) and $J_1=0.7$ (b,d) and $N=12$ (a,c). 
 In (c,d), the exponential decay of a single spin is shown for comparison.  
 The steady-state values in (a,b) are $\langle S_i^z\rangle_\mathrm{ss}= 1/N$.
\label{fig:s_z_time_evol_J_1_size}}
\end{figure}

Figure~\ref{fig:s_z_time_evol_J_1_size} shows the time evolution
$\langle S_\mathrm{edge}^z \rangle(t)$ for $L_z$ and $L_x$. Longer chains do not have a strong effect on the decay
time in the case of $L_x$, while they slow it down for $L_z$, see
Fig.~\ref{fig:s_z_time_evol_J_1_size}(b) and~(d). Note that the
steady-state values also depend on chain length in the $L_z$ case, see Eq.~\eqref{eq:s_z_steady_state}.
Time evolution for $L_x$ is dominated by $e^{-\gamma t/2}$, which represents the
exponential decay of  $\langle S_\mathrm{chain}^z\rangle(t)$, see
Eq.~\eqref{eq:s_z_chain_l_x}.

Figure~\ref{fig:s_z_time_evol_J_1_size}(c) discusses the impact of
$J_1$ on  $\braket{S_\mathrm{edge}^z}(t)$ for $L_x$. Very weakly coupled dimers with $J_1=0.1$ show a time evolution
$\mathrm{e}^{-\frac{\gamma t}{2}}$.
This can be explained by considering that a chain with $J_1=0.1$
consists of nearly noninteracting dimers, so that the edge states
resemble single spins. 
Larger values of $|J_1|$ lead to different decay dynamics that does,
however, not depend significantly on $J_1$. Chains from the topological dimer scenario, the
Haldane-like scenario, and also topologically trivial chains all
behave similarly. (Except for starting from different values at $t=0$.)
There is therefore no distinction in the $L_x$-driven time evolution of $\langle
S^z \rangle(t)$ between topological and topologically trivial chains.

In the case of  $L_z$ coupling, $J_1$ in contrast has an impact, see
Fig.~\ref{fig:s_z_time_evol_J_1_size}(a).   
Spin chains with nearly noninteracting dimers (e.g., $J_1=0.1$)
exhibit an extremely slow time evolution, whereas weaker dimerization
(e.g., $J_1=0.7$) gives a much faster time evolution.  
This is because $\langle S_\mathrm{chain}^z\rangle $ is conserved
under $L_z$ coupling, leading to $\langle S_i^z \rangle(t)
=\textrm{const.}$ in the limit of vanishing $J_1$.  
Hence, it makes sense that with decreasing $|J_1|$, the time evolution
towards the steady state becomes slower.

Finally, Fig.~\ref{fig:s_z_time_evol_J_1_size}(a) illustrates that chains
exhibiting weak dimer formation (e.g., $J_1 = 0.7$) and those
corresponding to the Haldane-like scenario show similar degrees of
robustness. For comparison, the topologically trivial chain with $J_1
= 1$ is also shown, and as expected, it reaches the steady-state value
more rapidly than the others.

\begin{figure}
  \includegraphics[width=\columnwidth]{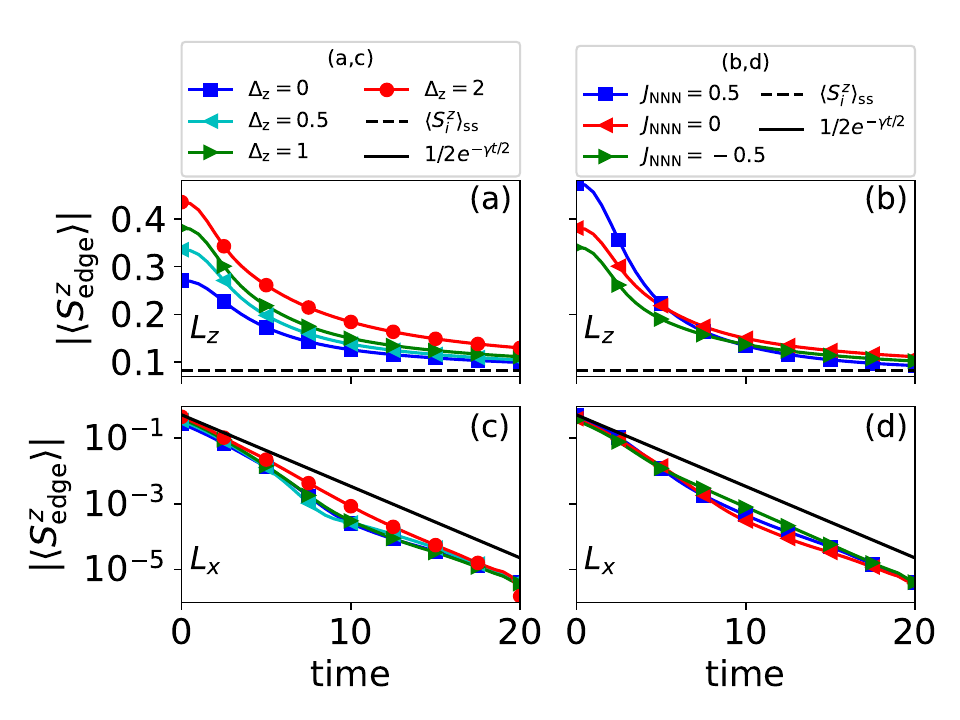}
    \caption{$\langle S_\mathrm{edge}^z\rangle(t)$ depending on NNN coupling and $\Delta_z$ for the triplet states with $\langle S_\mathrm{chain}^z\rangle(0)=\pm1$  for chains with $N=12$, $\gamma=1$,  $J_1=0.7$ and  $L_z$ (a,b) and  $L_x$ (c,d). 
In (c,d), the exponential decay of a single spin is shown for comparison.
The steady-state values in (a,b) are $\langle S_i^z\rangle_\mathrm{ss}= 1/N$.  
\label{fig:s_z_time_evol_anis_NNN}}
\end{figure}

Figure~\ref{fig:s_z_time_evol_anis_NNN}(a) illustrates that
$z$-anisotropy somewhat amplifies or reduces $\langle
S_\mathrm{edge}^z\rangle(t)$ on short-time scales under $L_z$
coupling. NNN coupling can amplify or rather reduce it, but only on
rather short time scales, see
Fig.~\ref{fig:s_z_time_evol_anis_NNN}(b).
For $L_x$ coupling, NNN coupling and $z$-anisotropy exert only a
negligible influence on $\langle S_\mathrm{edge}^z\rangle(t)$, see
Figs.~\ref{fig:s_z_time_evol_anis_NNN}(c, d). The primary factor
governing the time evolution for $L_x$ is the exponential decay
$\mathrm{e}^{-\frac{\gamma t}{2}}$, as seen above in Fig.~\ref{fig:s_z_time_evol_J_1_size}.

\begin{figure}
  \includegraphics[width=\columnwidth]{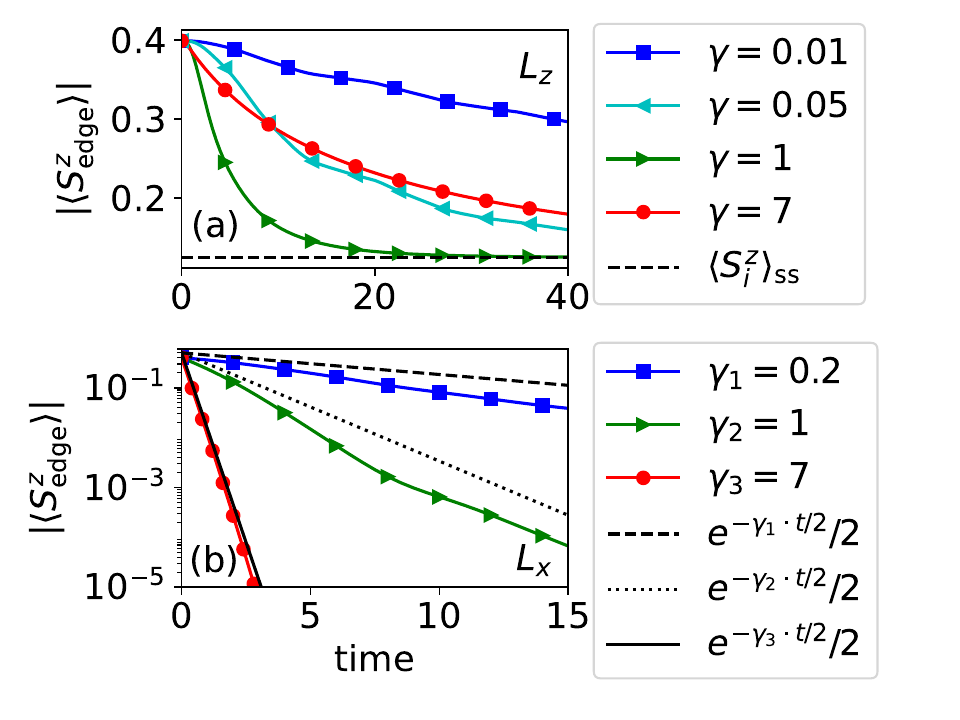}
    \caption{$\gamma$ dependency of $\langle S_\mathrm{edge}^z\rangle(t)$ for the triplet states with $\langle S_\mathrm{chain}^z\rangle(t=0)=\pm1$  for chains with $N=8$, $J_1=0.7$ and $L_z$ (a)   and $L_x$ (b). The steady-state value in (a) is $\langle S_i^z\rangle_\mathrm{ss}= 1/N$. In (b), the exponential decays of a single spin are shown for comparison.
     \label{fig:s_z_time_evol_gamma}}
\end{figure}

Time evolution under $L_x$ has so far been found to be nearly
independent of chain parameters, but it  strongly depends on $\gamma$,
see Fig.~\ref{fig:s_z_time_evol_gamma}(b).  In contrast, $L_z$ is
significantly influenced by some chain parameters ($J_1$ and $N$, less
for $\Delta_z$ and $J_\mathrm{NNN}$), and the impact of $\gamma$ is
less straightforward. As can be seen in
Fig.~\ref{fig:s_z_time_evol_gamma}(a), edge-state magnetization
exhibits a QZE for $L_z$. The minimum robustness  is typically
around  $\gamma \sim J_\mathrm{NN}$, with differences becoming more
pronounced as the disparity between  $J_1$ and $J_2$ increases.

The results presented in this section demonstrate that the time
evolution of the edge states in different chains under $L_x$ coupling
follows a trend represented by 
\begin{align}
\braket{S_\mathrm{edge}^z}(t) \sim \braket{S_\mathrm{edge}^z}(t=0)\  \mathrm{e}^{-\frac{\gamma t}{2}}.
\end{align} 
While this time evolution is not precisely $\propto \mathrm{e}^{-\frac{\gamma t}{2}}$, it is approximately so.
Remarkably, this time evolution pattern is consistent for both topological and topologically trivial chains and the same exponential decay behavior is observed for a single spin, as detailed in Appendix~\ref{sec:appendix_L_x}.
Consequently, based on $\braket{S_\mathrm{edge}^z}(t)$ for $L_x$,  it appears that there is no topological protection against environmental coupling.

 \subsection{Correlation function}\label{sec:correlation}
\begin{figure}
\includegraphics[width=\columnwidth]{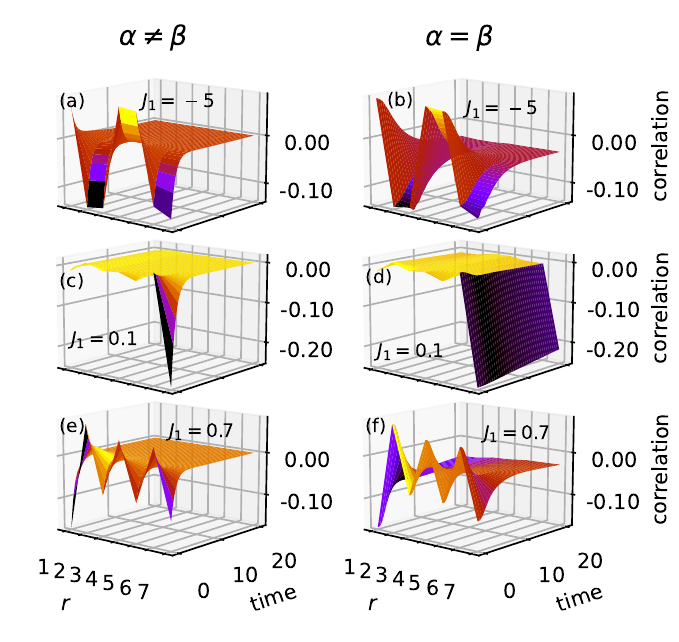}
\caption{Singlet state correlation $C(t,L_\beta,S_{r}^\alpha)$ Eq.~\eqref{eq:def_corr} with OBC for $N=8$, $J_\mathrm{NNN}=0$, $\Delta_\mathrm{z}=1$, $\gamma=1$ and $\alpha,\beta \in \{x,z\}$.
\label{fig:OBC_corr}}
\end{figure}

We then investigate the influence of the surface on the time evolution
of correlations
\begin{align}
C(t,L_\beta,S_{r}^\alpha)= \langle S_{1}^\alpha S_{1+r}^\alpha \rangle(t),
\label{eq:def_corr}
\end{align}
where  $\alpha,\beta \in \{x,z\}$, $S_1^\alpha$ is an edge spin and
$r$ denotes the distance between the spins. 
Notably, if the edge spins are in a singlet state, specific
combinations of jump and spin operators yield identical correlation
functions for $\Delta_z=1$, namely
$C(t,L_x,S_{i,j}^x)=C(t,L_z,S_{i,j}^z)$  and
$C(t,L_x,S_{i,j}^z)=C(t,L_z,S_{i,j}^x)$.

\begin{figure}
\includegraphics[width=\columnwidth]{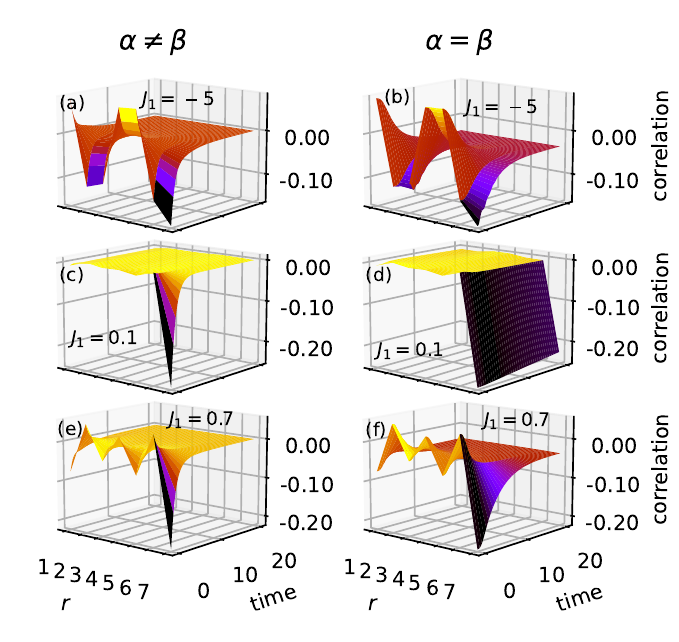}
\caption{Singlet state correlation $C(t,L_\beta,S_{r}^\alpha)$ Eq.~\eqref{eq:def_corr} with PBC for $N=8$, $J_\mathrm{NNN}=0$, $\Delta_\mathrm{z}=1$, $\gamma=1$ and $\alpha,\beta \in \{x,z\}$.
\label{fig:PBC_corr}}
\end{figure}

Correlations between spins diminish over time, except for finite-size
effects, as illustrated in Fig.~\ref{fig:OBC_corr} for OBC and
Fig.~\ref{fig:PBC_corr} for PBC. 
The difference between PBC and OBC is that for PBC $S_1$ and $S_N$ build a dimer, i.e., a singlet state, instead of free edge spins as in the case of OBC, and thus all spins are paired.
That explains why the largest correlation in Fig.~\ref{fig:PBC_corr}
is between $S_1$ and $S_N$.

These figures depict the correlations for the singlet state of edge
spins,  i.e., the state with the lowest energy.  
The steady-state correlations differ between two cases:
$C(t,L_\beta,S_{i,j}^\alpha)$ decreases to zero for $\alpha \neq
\beta$, while remaining finite for $\alpha=\beta$.  
While the correlation function $C(t, L_\beta, S_r^\alpha)$ with
$\alpha = \beta$ clearly indicates that edge states are more robust in
the dimerized scenario than in the Haldane-like scenario, for $\alpha
\neq \beta$, there is no distinction between the dimerized scenario,
the Haldane-like scenario, and topologically trivial chains.   
This can be seen in Fig.~\ref{fig:OBC_corr} by comparing the right and
left columns, and is reminiscent of the impact of $L_x$ discussed in the
previous section.

Some of the previously discussed results apply specifically to cases
where the edge spins form a singlet state, while others hold for
arbitrary states, including the triplet states of the edge spins. 
Specifically, the relations $C(t,L_x,S_{i,j}^x)=C(t,L_z,S_{i,j}^z)$
and $C(t,L_x,S_{i,j}^z)=C(t,L_z,S_{i,j}^x)$ hold solely for the
singlet state with $\Delta_z=1$.  
The correlation function for steady states is always zero if  $\alpha
\neq \beta$ in Eq.~\eqref{eq:def_corr}. 
A Finite-size effect where the steady state exhibits nonzero
correlation when $\alpha = \beta$ in Eq.~\eqref{eq:def_corr}, occurs
only for some states and parameters.

The previously discussed results hold even when $\gamma$ is varied.  
The occurrence of the QZE depends only on the relative orientation of
the jump operators and spin operators in the correlation function. The
QZE is always present for $\alpha = \beta$ and absent for $\alpha \neq
\beta$. When the QZE is observed, its minimum robustness typically
occurs around $\gamma \sim J_\mathrm{NN}$, with variations becoming
more significant as the difference between $J_1$ and $J_2$ increases.

\subsection{Entropy, fidelity and purity}\label{sec:entropy}
The time evolution of the density matrices, transitioning from the
initial pure eigenstates to typically mixed steady states, can be
analyzed using the following quantities: 
\begin{equation}
\begin{aligned}
S(t) &= - \mathrm{Tr}[ \rho(t) \mathrm{ln}(\rho(t))], \\
F(t) &=  \mathrm{Tr} [\rho(t=0)  \rho(t)], \\
P(t) &= \mathrm{Tr}[\rho^2(t)],
\end{aligned}
\label{eq:def_S_F_P}
\end{equation}
where $S$ is the von Neumann entropy, $F$ is the Uhlmann fidelity and $P$ the purity. 

\begin{figure}
  \includegraphics[width=\columnwidth]{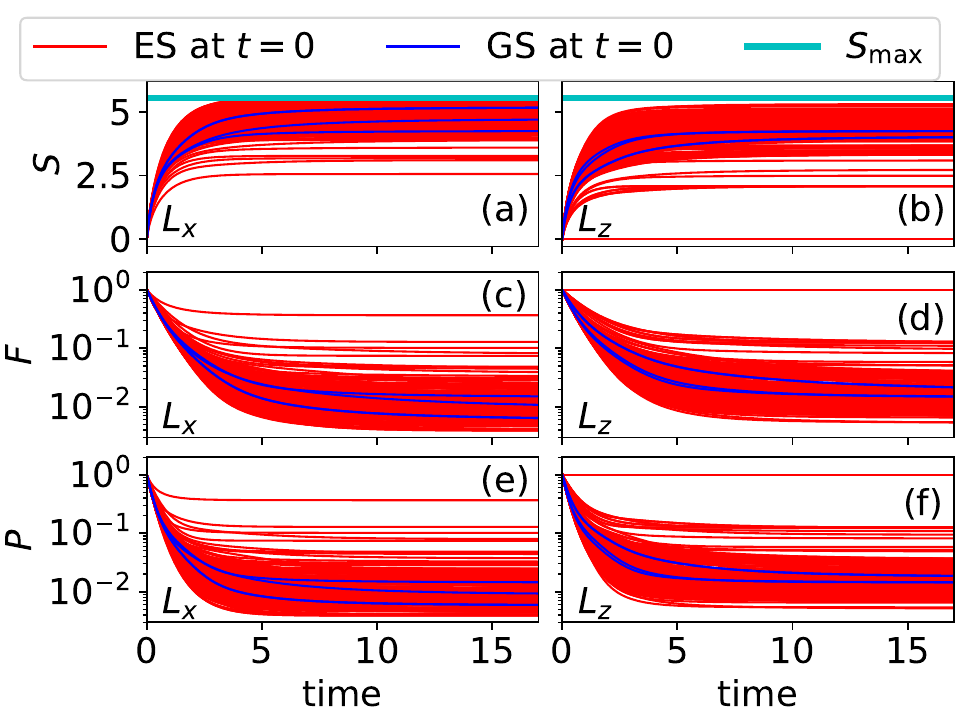}
    \caption{ Entropy ($S$), fidelity ($F$) and purity ($P$) for a chain with $N=8$, $\gamma=1$,  $J_1=0.7$ and with  $L_x$, in (a,c,e) and $L_z$, in (b,d,f).
 Similar time evolution occurs for both the four states of the topological ground state (GS) and the excited states (ES) of the closed chain.
\label{fig:J_1__0_7__S_F_Tr}}
\end{figure}

The time evolution of the quantities defined in Eq.~\eqref{eq:def_S_F_P} exhibits no significant difference between the topological ground states and the excited states, as shown in Fig.~\ref{fig:J_1__0_7__S_F_Tr}.
In addition, the time evolution shows no significant discrepancy between 
$L_x$ and $L_z$, except for those eigenstates of the Hamiltonian that 
are also steady states. 
Since the total magnetization $\langle S_\mathrm{chain}^z \rangle$ is
a conserved quantity for
$L_z$~\cite{PhysRevA.97.052106,PhysRevE.92.042143}, these  two
eigenstates of the Hamiltonian that have $\langle S_\mathrm{chain}^z
\rangle =\pm\frac{N}{2}$ are also steady states. They are thus
unaffected by $L_z$ and give the horizontal lines in Fig.~\ref{fig:J_1__0_7__S_F_Tr}.
As these states represent topologically trivial ferromagnetic states,
no stable topological states exist.

For $L_x$, the states with the highest entropy reach the maximal 
possible entropy value $ S_\mathrm{max} = \ln(\dim(H)) $,
where \( \dim(H) \) represents the dimension of the Hilbert space. In
contrast, for $L_z$, the largest achievable entropy values are always
below $S_\mathrm{max}$. This is illustrated in
Figs.~\ref{fig:J_1__0_7__S_F_Tr}(a) and \ref{fig:J_1__0_7__S_F_Tr}(b),
where $S_\mathrm{max}$ is shown.

The fit-function
\begin{align}
f(t) = at^{-b} e^{-ct}+d,
\label{eq:fit_S_F_Tr}
\end{align}
is applicable to all quantities defined in Eq.~\eqref{eq:def_S_F_P},
revealing consistent patterns in time evolution across various chains. 
These functions, characterized by fit parameters $a$, $b$, $c$, and
$d$, are inspired by analytical calculations of the fidelity for
different systems~\cite{PhysRevLett.122.040604,2208.07732}. 
However, when $\gamma \gg J_\mathrm{NN}$, the fit-function in
Eq.~\eqref{eq:fit_S_F_Tr} is no longer applicable. 
 This behavior is consistent with the assumption of $\gamma \ll
 J_\mathrm{NN}$ employed in the analytical
 derivations~\cite{PhysRevLett.122.040604,2208.07732}. 
Additionally, the fit functions do not apply to states remaining pure,
such as the horizontal lines in Figs.~\ref{fig:J_1__0_7__S_F_Tr}(b),
\ref{fig:J_1__0_7__S_F_Tr}(d) and \ref{fig:J_1__0_7__S_F_Tr}(f). Aside
from these exceptions, fit errors are negligible.

The influence of chain parameters on entropy, fidelity, and purity is minimal. 
Although fitting parameters  rely on chain-specific properties, their
dependence on chain parameters does not reveal a discernible trend
regarding the dominance of algebraic or exponential decay.  
Fit functions remain universally valid for chains
across the Haldane-like scenario, the topological dimer scenario, and
for topologically trivial chains.  
Consequently, there seems to be no distinction in entropy, fidelity,
and purity between topological and topologically trivial chains.  
On the other hand, the lifetime determined by $\Theta_\mathrm{PT}(t)$ exhibits significant variability
among different parameters of the Hamiltonian, as discussed in
Sec.~\ref{sec:phase_diagram}.  
Fidelity, purity, and entropy are nearly identical across chains with
vastly different $\Theta_\mathrm{PT}(t)$.

The occurrence of the QZE in entropy, purity, and fidelity depends on both the choice of the jump operator and the specific parameters of the Hamiltonian. Additionally, even when the QZE occurs, some states may exhibit it while others may not.
When the QZE is observed, its weakest robustness usually occurs around $\gamma \sim J_\mathrm{NN}$, with variations becoming more noticeable as the difference between $J_1$ and $J_2$ increases.

\subsection{Spectral gap}\label{sec:spectral_gap}

The long-term behavior can be described by relaxation times. 
For the LME, a relaxation time can be calculated based on the
eigenvalues $\lambda_\alpha$ of $\mathcal{L}$, which are present in
the exponential function of~\eqref{eq:LME_vec_solution}. 
The real parts $\mathrm{Re}(\lambda_\alpha)$ indicate how quickly the
steady state is reached. 
The slowest exponential decay, determining the long-term time
evolution, can be used to define  
\begin{align}
\frac{1}{\tau}= \Delta\coloneqq - \max\limits_{\substack{\alpha
    \\ \text{Re}[\lambda_\alpha]\neq 0}}{\mathrm{Re}(\lambda_\alpha)},  
\label{eq:def_spectral_gap}
\end{align}
where the inverse of the spectral gap $\Delta$ is the relaxation time
$\tau$~\cite{PhysRevB.99.174303,PhysRevLett.111.150403,PhysRevA.89.022118,PhysRevE.92.042143,PhysRevLett.132.070402}.

\begin{figure}
  \includegraphics[width=1.1\columnwidth]{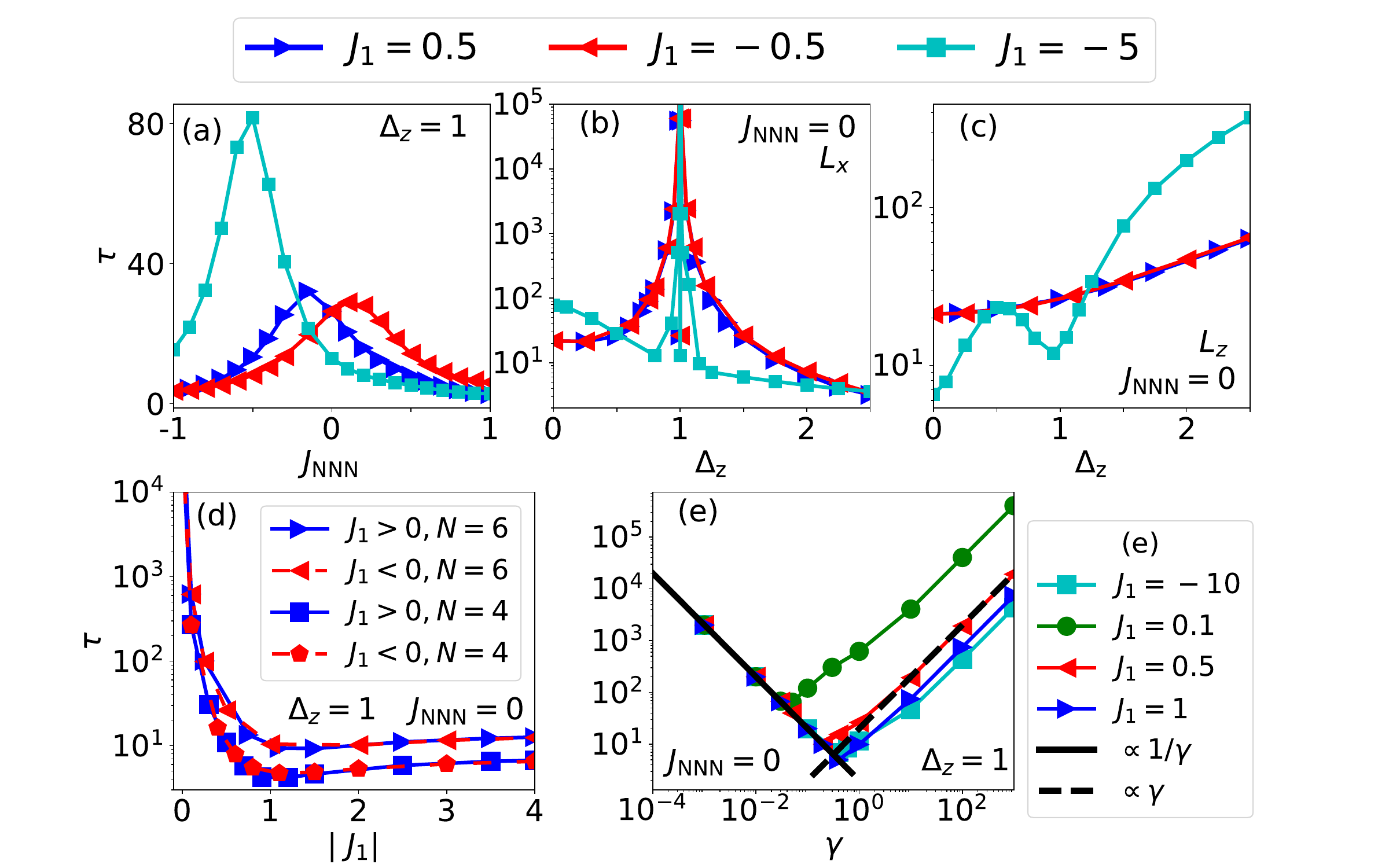}
   \caption{Inverse spectral gap $\tau$ Eq.~\eqref{eq:def_spectral_gap} depending on the parameters of the chain. There is always $\gamma=1$ except in (e), there is always $N=6$ except in (d). There is $\Delta_z=1$ and $J_\mathrm{NNN}=0$ except for the cases where they are explicitly varied.
    (a), (d) and (e)  are the same values for $L_x$ and $L_z$ and there is  $L_x$ in (b) and  $L_z$ in (c). 
\label{fig:spectral_gap_NNN_z_anis}}
\end{figure}

$\tau$ does not provide information regarding topological properties,
such as the robustness of edge states, as these features vanish before
reaching long-term time evolution, as discussed in the previous
sections.  Moreover, this is also evident in
Fig.~\ref{fig:spectral_gap_NNN_z_anis}(d), which shows $\tau$ as a
function of $J_1$. The difference between FM and AFM $J_1$ is merely
a consequence of finite-size effects in isotropic chains lacking NNN
coupling. Therefore, $\tau$ is identical for both the Haldane-like
scenario ($J_1<-1$) and the dimer scenario without edge states
($J_1>1$).

Comparing the results for $N=4$ and $N=6$ in
Fig.~\ref{fig:spectral_gap_NNN_z_anis}(d) reveals that larger spin
chains exhibit greater $\tau$.  Apart from this, $\tau$ remains quite
similar for different chain lengths. 
$\tau$ is further identical for both jump operators $L_x$ and $L_z$ in isotropic chains.
Therefore, the results depicted in
Figs.~\ref{fig:spectral_gap_NNN_z_anis}(a),
\ref{fig:spectral_gap_NNN_z_anis}(d) and
\ref{fig:spectral_gap_NNN_z_anis}(e) apply to both $L_x$ and $L_z$.

However, for anisotropic chains  as in
Figs.~\ref{fig:spectral_gap_NNN_z_anis}(b) and
\ref{fig:spectral_gap_NNN_z_anis}(c),  differences emerge between the
two jump operators.  
Figures~\ref{fig:spectral_gap_NNN_z_anis}(b) and
\ref{fig:spectral_gap_NNN_z_anis}(c) highlight that tiny deviations
from the isotropic case with $L_x$ can cause a significant change in
$\tau$, whereas  there is no drastic change for $L_z$. 
This difference between the jump operators stems from the alteration
in the degeneracy of the steady state,  as discussed in
Sec.~\ref{sec:S_z_exp_val}.  
The $z$-anisotropy can either increase or
decrease $\tau$ compared to the isotropic chain, depending on the jump
operator and the chain parameters. This is seen
in Figs.~\ref{fig:spectral_gap_NNN_z_anis}(b) and (c).
The independence of the sign of $J_1$ (except the finite-size effect,
see Fig.~\ref{fig:spectral_gap_NNN_z_anis}(d)) can be extended from
isotropic chains to anisotropic chains.

The influence of NNN coupling on $\tau$ is shown in Fig.~\ref{fig:spectral_gap_NNN_z_anis}(a).
In chains from the topological dimer scenario without frustration, $\tau$ reaches its maximum value.
Conversely, for chains from the Haldane-like scenario, the maxima
occur for the sign of $J_\mathrm{NNN}$ that leads to frustration
between NN and NNN coupling. 
NNN coupling can enhance the robustness compared to
$J_\mathrm{NNN}=0$.

The results presented thus far are obtained for $\gamma = 1$, leaving
the question of  how the results are affected by $\gamma$. 
Regardless of $\gamma$, the independence of $\tau$ on the jump
operator and the sign of $J_1$ for isotropic chains lacking NNN
coupling remains consistent.  
The influence of $\gamma$ on
Figs.~\ref{fig:spectral_gap_NNN_z_anis}(a)-\ref{fig:spectral_gap_NNN_z_anis}(d)
primarily can mostly be captured by rescaling the $\tau$-axis, with
only occasional additional variations. 
For instance, decreasing $\gamma$ sharpens the peaks in
Fig.~\ref{fig:spectral_gap_NNN_z_anis}(a) leading to a $\tau$  nearly
independent of $J_\mathrm{NNN}$, except for peaks occurring nearly at
the same position as those in Fig.~\ref{fig:spectral_gap_NNN_z_anis}(a).

The QZE consistently occurs for $\tau$, as shown in
Fig.~\ref{fig:spectral_gap_NNN_z_anis}(e). This is evident from the
behavior of $\tau$, which decreases as $\gamma$ increases but begins
to rise again at larger $\gamma$. 
When the QZE is observed, its robustness is weakest around $\gamma
\sim J_\mathrm{NN}$, with variations becoming more pronounced as the
disparity between $J_1$ and $J_2$ increases.

In Fig.~\ref{fig:spectral_gap_NNN_z_anis}(e), two lines that are
proportional to  $\gamma$ and $1/\gamma$ are shown, representing the
behavior observed for a single spin (for $L_x$) and two spins (for
$L_z$), as depicted
Fig.~\ref{fig:spectral_gap_analytical_calculation}. 
The figure clearly illustrates that the results for the larger chains
exhibit similar  $\gamma$ dependency as a single spin with $L_x$
(respectively two spins for $L_z$).  
Thus, this $\gamma$ dependency  is independent of the number of spins
and is not a consequence of the topological properties.

\section{Discussion and Conclusions}\label{sec:discuss}

We have investigated the robustness of edge states in topological spin chains on surfaces against surface coupling. 
To accomplish this, we employed the LME to analyze the time evolution of several parameters, including the ground-state degeneracy, edge-state magnetization, entropy, correlation functions, and spectral gap. 
Our study primarily concentrated on rather short chains, as the
experimental implementation of the system under consideration involves
spin chains on surfaces, where the topological edge states can be
measured with an STM~\cite{2403.14145, zhao2024tunable}, and
achievable chain lengths are comparable.

Despite notable finite-size effects and increasing robustness with
increasing number of spins, short chains still demonstrate parameter
ranges with a near degeneracy of edge states. 
We extend here our previous
result~\cite{sattler2024topologically}, which showed that approximate ground-state
degeneracy is more robust in dimerized chains than in Haldane-like chains, from closed
chains to open chains. 

However, steady states are topologically trivial, making spin-spin
correlations vulnerable to environmental coupling. These correlations,
including edge and bulk spins, decay over time (except finite-size
effects) in both topological and trivial chains, showing the lack of
topological protection. This reflects the expected loss of
entanglement due to environmental influences~\cite{RevModPhys.81.865}
and is here studied via several quantities.

Some of the criteria examined depend on the topological properties, others do not. 
For instance, the spectral gap depends on chain parameters, but not on
topological properties. It
is given by the system's long-term time evolution, while topological
features characterizing the initial near ground-state degeneracy are
typically lost over short timescales.

Similarly, correlation functions $C(t, L_\beta, S_r^\alpha)$
exhibit similar decay for both topological and topologically trivial
chains when $\alpha \neq \beta$. However, when $\alpha = \beta$,
dimerized chains continue to have significantly more robust edge states
compared to Haldane-like chains. Edge-state magnetization $\langle S_\mathrm{edge}^z\rangle(t)$ for
$L_z$ is also more robust in the dimerized scenario than in
the Haldane-like scenario. In contrast, its time evolution under $L_x$
is dominated by exponential decay in all cases, without relevant differences
between topological and topologically trivial chains. 

Topological edge spins thus turn out to resemble single spins: 
When the jump operator is  $L_x$, $\langle S_i^z \rangle(t)$ decays with nearly the same 
exponential decay for both topological edge spins and single spins. 
Furthermore, the exponential decay of the entire chain $\langle
S_\mathrm{chain}^z \rangle(t)$ mirrors that of a single spin
precisely.

Previous research has investigated the AKLT
model~\cite{PhysRevLett.59.799} coupled to a quantum
bath~\cite{PhysRevB.104.094306,
  PhysRevLett.127.086801,PhysRevLett.126.237201}.  
These studies have demonstrated the destabilization of edge states
within the AKLT chain when subjected to a quantum bath, corroborating
the findings of our study.  
Additionally, an increase in entropy within our model aligns with
observations made in prior research on the AKLT
model~\cite{PhysRevB.104.094306,PhysRevLett.127.086801}.

The time evolution of fidelity, purity, and entropy remains basically unaffected
by the parameters of the chain, for  topological and topologically
trivial chains.  
The Uhlmann fidelity, exhibiting a mix of exponential and algebraic
decay, shows universality across various systems. This has been
observed in systems like chains with noisy magnetic fields, a 2D Bose
gas in the superfluid phase with localized particle
heating~\cite{PhysRevLett.122.040604}, and the 2D Kitaev honeycomb
model~\cite{2208.07732}.  
Previous studies focused only on fidelity, while this work
demonstrates that the same function also describes entropy and
purity. Additionally, while prior
research~\cite{PhysRevLett.122.040604,2208.07732} concentrated on
dephasing $S^z$, this study extends the analysis to $S^x$ jump
operators, showing similar time evolution.  
This suggests that the Uhlmann fidelity has a common feature that
holds true regardless of the system's dimensions, Hamiltonian, jump
operator, or topology. However, it's important to note that this
behavior doesn't always apply~\cite{PhysRevB.109.L041107}.

Varying $\gamma$ shows that a
QZE~\cite{breuer_petruccione_book,10.1063/1.523304} can occur. 
The QZE has been theoretically discussed in various open
systems~\cite{PhysRevB.101.075139,PhysRevLett.122.040402,PhysRevB.109.085115,PhysRevA.94.043609,PhysRevA.106.013707,PhysRevB.99.174303,PhysRevLett.126.110404,PhysRevLett.115.083601}. It
has furthermore been confirmed through experimental
observations~\cite{PhysRevLett.87.040402,10.1126/science.1155309,PhysRevA.41.2295,PhysRevLett.115.140402}. 
For a specific set of chain parameters, a QZE can be seen in one 
criterion,  while being absent for another. The spectral
gap consistently displays a QZE. The presence 
of the QZE for $\langle S_\mathrm{edge}^z\rangle(t)$ depends on the
jump operator and for correlation functions on the relative orientation
between the jump operators and the spin operators, but it is
independent of the chain parameters. In 
contrast, observations of the QZE for $\Theta_\mathrm{PT}$, the
entropy, the  purity, and the fidelity depend on all the parameters
of the chain and the jump operators. However, a common feature in
all cases where the QZE is seen, is that the weakest robustness 
typically arises around $\gamma \sim J_\mathrm{NN}$.

The investigation of topological edge states and their detection via
STM~\cite{RevModPhys.81.1495,RevModPhys.91.041001,GAUYACQ201263} is
of wide interest, extending beyond the specific focus on spin chains
explored in this study.  
Another instance in one dimension involves the identification of edge
states in a Kitaev chain~\cite{Rev_edges_STM_21}.  
However, edge states and their potential signatures in STM
measurements are also relevant in higher dimensions, e.g.,
for the Kitaev honeycomb
model~\cite{PhysRevB.102.134423,PhysRevLett.126.127201,PhysRevLett.125.267206}. 
Progress in this direction includes the experimental realization of
monolayers of a Kitaev spin liquid candidate on a
surface~\cite{D2NR02827A,2403.16553}.  
Additionally, an alternative approach for realizing the Kitaev
honeycomb model includes employing quantum
dots~\cite{cookmeyer2023engineering}.

\begin{acknowledgments}
The authors acknowledge support by the state of Baden-Württemberg through bwHPC.
\end{acknowledgments}

\section{References}

\bibliography{open_chain}

\newpage

\appendix

\section{Numerics}\label{appendix:LME}
The Lindblad operator $\mathcal{L}$, see Eq.~\eqref{eq:linbdlad_vec_LME}, is a so-called superoperator, since it acts on a matrix. 
Performing calculations involving superoperators presents significant challenges.
Nevertheless, it is feasible to reframe LME as an eigenvalue problem by leveraging the  Choi–Jamiolkowski isomorphism~\cite{CHOI1975285,JAMIOLKOWSKI1972275}.
This method, alternatively termed vectorization, involves arranging the columns of the density matrix $\rho$ into a vector.
The vectorized~\cite{10.1088/1751-8121/aae4d1,PhysRevB.99.174303,1510.08634}  form of the LME Eq.~\eqref{eq:lindblad_LME} is 
\begin{equation}
\begin{aligned}
&\frac{\mathrm{d}}{\mathrm{dt}}  \vert\rho\rangle\!\rangle =    \mathcal{L}  \vert\rho\rangle\!\rangle = \biggl[ - \mathrm{i} \left( \mathbb{1} \otimes H - H^\mathrm{T} \otimes \mathbb{1} \right)\\
 &+  \gamma \sum_i  \left( {\left(L_i^\dagger \right)}^\mathrm{T} \otimes L_i^{\vphantom{\dagger}} - \frac{1}{2} \left(  \mathbb{1} \otimes  L_i^\dagger  L_i^{\vphantom{\dagger}} +   {\left(L_i^\dagger  L_i^{\vphantom{\dagger}} \right)}^\mathrm{T} \otimes \mathbb{1} \right) \right)  \biggr]  \vert\rho\rangle\!\rangle.
\end{aligned}
\label{eq:linbdlad_vec_LME}
\end{equation}
Here, $\mathcal{L}$ represents the  $n^2 \times n^2 $ Lindblad matrix, $H$ the $n \times n $ Hamiltonian of the system, $\mathbb{1}$ denotes $n \times n $ identity matrix, and $\vert\rho\rangle\!\rangle$ signifies the density matrix in vectorized form.
This differential equation can be readily solved using an exponential ansatz
\begin{subequations} 
\label{eq:LME_vec_solution} 
\begin{align}
 \vert\rho(t)\rangle\!\rangle =  \mathrm{e}^{\mathcal{L}t} \vert\rho(0)\rangle\!\rangle &=  S \mathrm{e}^{\Lambda t} S^{-1}  \vert\rho(0)\rangle\!\rangle \label{eq:LME_vec_solution_exact}\\ 
			& = \left[ \sum_{m=0}^{\infty} \frac{1}{m!} \mathcal{L}^mt^m\right]  \vert\rho(0)\rangle\!\rangle, \label{eq:LME_vec_solution_power_series}
\end{align} 
\end{subequations} 
where we adopt the initial condition that the system starts with the density matrix of the closed system at $t=0$.
For full diagonalization (FD) of small chains up to $N=6$, we  utilize Eq.~\eqref{eq:LME_vec_solution_exact}.
In this equation, the matrix $S$ contains the eigenvectors of $\mathcal{L}$ and the diagonal elements of the  matrix $\Lambda$ contain the eigenvalues of $\mathcal{L}$.

However, performing FD entails significant computational effort.
Therefore, an alternative approach is to utilize  Eq.~\eqref{eq:LME_vec_solution_power_series} to compute an approximate solution for longer chains.
This method involves selecting a time step $\Delta t$ and computing the power series at $t=0$ to derive $\vert\rho(\Delta t)\rangle\!\rangle$. 
Subsequently, we iteratively calculate the power series to obtain  $\vert\rho(2 \Delta t)\rangle\!\rangle$ and so forth.
A good parameter selection is an expansion up to the $m_\mathrm{max}=10$ order and a time step $\Delta t=0.1$.
While in some cases, parameters such as $m_\mathrm{max}=5$ and $\Delta t =0.5$ may be adequate, they are not universally applicable.

The numerical computation of the LME, see Eq.~\eqref{eq:LME_vec_solution}, is an eigenvalue problem with a dimension squared relative to the dimension of the Hamiltonian.
Consequently, the diagonalization of the Hamiltonian is significantly faster compared to the computation of  $\vert\rho(t)\rangle\!\rangle$, leading to the utilization of FD for the Hamiltonian.

\section{One and two spins}

In the main section, we conducted comparisons between the outcomes obtained from a chain and those from a single spin. 
Here, we derive the single- and two-spin results used for comparison in the main text.

To simplify the calculation, we use a basis in which the  Hamiltonian is diagonal
\begin{align}
H=
\left(
\begin{array}{rr}
E_1 & 0 \\
0  & E_2
\end{array}
\right),
\label{eq:appendix:ham}
\end{align}
with the energy difference denoted as $\Delta E = E_1-E_2$.
The next sections briefly discuss the application of the LME to this Hamiltonian. 
First, in Sec.~\ref{sec:appendix_L_z}, we examine the jump operator $L=S^z$, followed by a discussion in Sec.~\ref{sec:appendix_L_x} on $L=S^x$ and conclude in Sec.~\ref{sec:appendix_S_+_S_-} by examining  alternative jump operators in place of $L = S^x$.

\begin{figure}
  \includegraphics[width=\columnwidth]{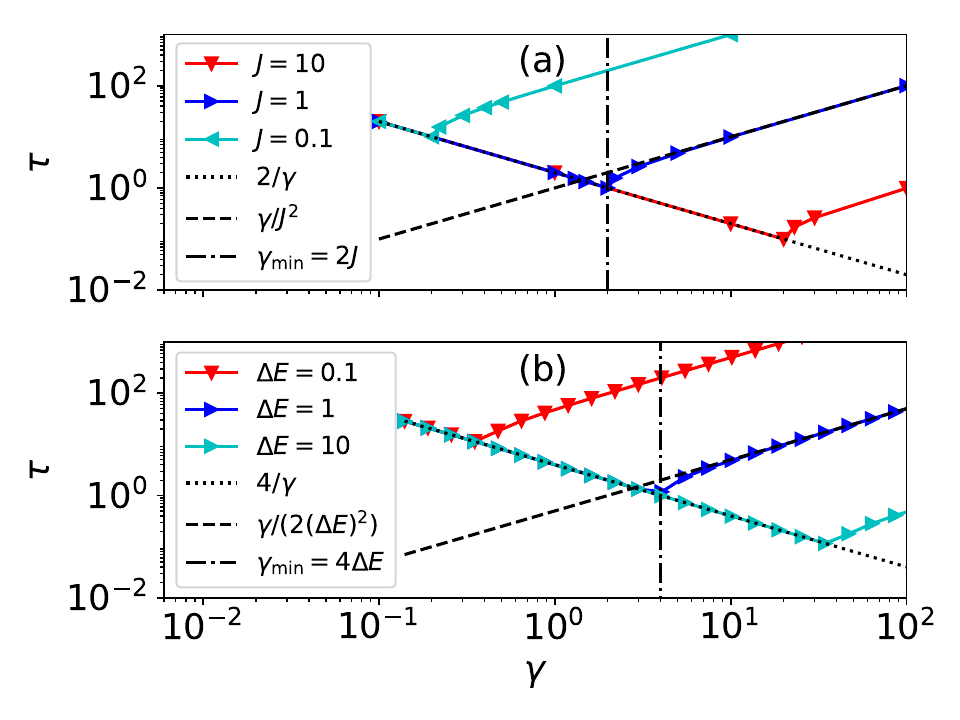}
   \caption{The inverse spectral gap $\tau$, as defined in Eq.~\eqref{eq:def_spectral_gap}, is depicted for two spins with $L_z$ (numerical calculation) in (a) and for a single spin with $L_x$ (analytical calculation) in (b). }
\label{fig:spectral_gap_analytical_calculation}
\end{figure}

\subsection{Jump operator $L=S^z$}\label{sec:appendix_L_z}
Following a straightforward calculation of the LME in vectorized form, see Eq.~\eqref{eq:linbdlad_vec_LME}, for the diagonal Hamiltonian Eq.~\eqref{eq:appendix:ham} and the jump operator $L=S^z$,  we obtain
\begin{align}
\rho(t) = \left(
\begin{array}{cc}
\rho_{11}(0) & \rho_{12}(0) \mathrm{e}^{-\mathrm{i} t \Delta E } \mathrm{e}^{-t\frac{\gamma}{2}}\\
\rho_{21}(0) \mathrm{e}^{\mathrm{i} t \Delta E } \mathrm{e}^{-t\frac{\gamma}{2}} & \rho_{22}(0) \\
\end{array}
\right).
\end{align}
The diagonal elements remain constant, reflecting the conservation of magnetization  $\langle S^z \rangle (t)= \mathrm{const}$, a characteristic inherent of the jump operator $L=S^z$~\cite{PhysRevA.97.052106,PhysRevE.92.042143}.
Additionally, the off-diagonal elements decay, leading to a diagonal steady-state density matrix, highlighting the role of $L = S^z$ in inducing decoherence.

The inverse of the spectral gap, as defined in Eq.~\eqref{eq:def_spectral_gap}, is determined to be 
\begin{align}
\tau = \frac{2}{\gamma}. 
\end{align}
It becomes evident that there is no QZE in the presence of decoherence for a single spin.
However, a system consisting of two spins with Heisenberg coupling $J$ and the jump operators $L_1=S_1^z$ and $L_2=S_2^z$, does exhibit the QZE within the spectral gap, as depicted in  Fig.~\ref{fig:spectral_gap_analytical_calculation}(a).
Thus, a minimum system size of two spins is necessary for the presence of a QZE due to decoherence.
Notably, a single spin can also demonstrate the QZE if the jump operator is $L=S^x$, as discussed in Appendix~\ref{sec:appendix_L_x} and depicted in Fig.~\ref{fig:spectral_gap_analytical_calculation}(b).

\subsection{Jump operator $L=S^x$}\label{sec:appendix_L_x}
Similar to the discussion of the jump operator $L=S^z$ in Sec.~\ref{sec:appendix_L_z}, we now discuss the jump operator $L=S^x$ for the Hamiltonian Eq.~\eqref{eq:appendix:ham}.
The diagonal elements of $\rho(t)$ are
\begin{align}
\rho_{11}(t) &= \frac{1}{2} - \frac{1}{2} \left( 1 - 2 \rho_{11}(0) \right)   \mathrm{e}^{\frac{-\gamma t}{2}} \\
\rho_{22}(t) &= \frac{1}{2}  - \frac{1}{2} \left(1 - 2 \rho_{22}(0) \right)   \mathrm{e}^{\frac{-\gamma t}{2}}
\end{align}
and the steady state is
\begin{align}
\rho_{\mathrm{ss}}= 
\left(
\begin{array}{cc}                                
\frac{1}{2} & 0  \\ 
0 & \frac{1}{2}  \\ 
\end{array}
\right). 
\end{align}
This leads to 
\begin{align}
\braket{S^\mathrm{z}}(t) = \frac{1}{2} \left( 
\rho_{11}(0) - \rho_{22}(0) 
\right)
\mathrm{e}^{\frac{-\gamma t}{2}}
\label{eq:analytical_S_x_S_z_chain} 
\end{align}
and
\begin{align}
\braket{S^\mathrm{z}}_\mathrm{ss}=0.
\end{align}
This result is in perfect agreement with the findings obtained from calculations conducted on longer chains, as discussed in  Sec.~\ref{sec:S_z_exp_val}. 

The spectral gap, see Eq.~\eqref{eq:def_spectral_gap}, is shown in  Fig.~\ref{fig:spectral_gap_analytical_calculation}(b).
The spectral gap exhibits a QZE, suggesting its occurrence does not require topology or complex interactions. 

\subsection{Jump operators $L_1=S_+$ and $L_2=S_-$}\label{sec:appendix_S_+_S_-}
Instead of employing $L=S^x$ as a jump operator, an alternative approach involves utilizing two distinct jump operators $L_1=S_+$ and $L_2=S_-$, each with the dissipation strengths $\gamma_+$ and $\gamma_-$, respectively.

In contrast to $L=S^x$, the combination of these two jump operators does not mix the off-diagonal elements of $\rho$.
This significantly impacts the time evolution, e.g., $\tau$ does not exhibit the QZE.
Consequently, using the two jump operators, $L_1 = S_+ $ and $L_2 = S_-$, separately has a different physical meaning than using just one jump operator, $L = S^x$.
This distinction is elaborated in Sec.~\ref{section:model}, and a similar discussion regarding the meaning of jump operators is found in a study concerning decoherence in a three-level system involving nitrogen vacancies~\cite{PhysRevB.94.134107}.

\end{document}